# Recent Activities of a European Union Joint Research Project on Metrology for Emerging Wireless Standards

Tian Hong Loh, *Member, AMTA.*
National Physical Laboratory (NPL)
Teddington, United Kingdom
tian.loh@npl.co.uk

Wei Fan, *Senior Member, AMTA.*
Aalborg University
Aalborg, Denmark
wfa@es.aau.dk

Akram Alomainy
Queen Mary University of London (QMUL)
London, United Kingdom
a.alomainy@qmul.ac.uk

Tas Emrah,
Frédéric Pythoud
Federal Institute of Metrology METAS
Berne-Wabern, Switzerland
emrah.tas@metas.ch
frederic.pythoud@metas.ch

Djamel Allal
Laboratoire National de Métrologie et d'Essais (LNE)
Paris, France
djamel.allal@lne.fr

*Abstract*—Emerging wireless technologies with Gbps connectivity, such as the 5th generation (5G) and 6th generation (6G) of mobile networks, require improved and substantiating documentation for the wireless standards concerning the radio signals, systems, transmission environments used, and the radio frequency exposures created. Current challenges faced by the telecommunications sector include the lack of accurate, fast, low-cost, and traceable methods for manufacturers to demonstrate 5G/6G product verifications matching customers' specifications. This paper gives an update on the recent research and development activities from an EU Joint Research Project entitled "metrology for emerging wireless standards" (MEWS) in support of the above.

## I. Introduction

The digital economy is essential for global wealth creation and it is increasingly underpinning all aspects of social and business activities. Information handling services (IHS) economics have estimated that the 5th generation (5G) networks and beyond will enable USD12.3 trillion of global economic output in 2035 [1]. The growing demand for high-speed communication has driven the need to foster innovation in emerging wireless technologies to support 5G, 6th generation (6G), and automotive wireless communications. The immediate challenge for communication networks is to scale up to meet increasing traffic requirements, as well as serving new elaborate machine-to-machine systems and applications. Metrology has a pivotal role to ensure the product quality and the end-user confidence, and ultimately to improve the competitiveness of the telecommunications industry [2].

With the industrial exploitation in emerging wireless systems and adoption of complex NR signals and large-scale multi-antenna technologies, such as multi-user multiple-input-multiple-output (MU-MIMO) and massive MIMO at different radio frequency (RF) bands, such as sub-6GHz, millimetre-wave (mm-wave) and sub-THz bands, verification of emerging NR product, especially, for high-volume beam-reconfigurable products, have become very time consuming and involve complicated procedures and equipment, leading to high cost [2, 3]. While the 5G standardisation and the 6G definition processes are ongoing, the key challenges are a lack of practical metrology to support NR over-the-air (OTA) testing, sub-THz channel characterisation, and RF exposure assessment. Several multinational industries, research communities and standards bodies (e.g. 3GPP (Third-Generation Partnership Project) [4], ETSI (European Telecommunication Standards Institute) [5] and IEEE (Institute of Electrical and Electronics Engineers) [6]), are now actively seeking for the improved process control on NR OTA methods and RF exposure assessment as well as the empirical sub-THz channel characterisation to support 6G definitions, research and development. Furthermore, ETSI has set millimetre wave transmission (mWT) and THz spectrum with wider channel bandwidth as its standardisation priorities [7] whereby ITU-R (International Telecommunication Union Radiocommunication Sector) highlights in [8] that active research in this area would require extensive standardisation efforts.

In support of the above, a three-year Joint Research Project entitled "metrology for emerging wireless standards" (MEWS) [9] has been launched since October 2022 in the framework of the new European Association of National Metrology Institutes (EURAMET) European Partnership on Metrology programme. The planned developments concern three aspects, namely: 1) NR OTA metrology; 2) sub-THz radio propagation metrology; 3) NR exposure metrology. This project involves 17 partners including 5 European national measurement institutes (NMIs) together with the academy and global industrial manufacturers. In this paper, an update on the recent activities from this EU Joint Research Project is presented.

## II. NEW RADIO OVER-THE-AIR METROLOGY

This section summarizes the activities related to the development and validation of practical, fast, and cost-effective radiated OTA performance and conformance testing methods for 5G NR (including mobile handsets, base stations and automotive systems) operating at sub-6GHz and mm-wave bands. There are two main tasks: 1) development of efficient OTA performance testing methods; and 2) development of efficient OTA RF parametric measurement solutions.

For the OTA performance testing method development, the key objectives are to: 1) extend wireless cable solution, which supports sub-6GHz 2 × 2 MIMO radios in the current state-of-art, to support sub-6GHz high-order MIMO radios and mm-wave 2×2 MIMO radios; 2) improve the efficiency and performance of the test zone validation algorithm for the multi-probe anechoic chamber (MPAC) solution at mm-wave bands; and 3) improve the cost-efficiency and accuracy of the reverberation chamber plus channel emulator (RC+CE) for the sub-6GHz and mm-wave-bands.

For the OTA RF parametric measurement solution development, the key objectives are to: 1) develop fast OTA conformance testing methods for NR RF metric, e.g. equivalent isotropic radiated power (EIRP), effective isotropic sensitivity (EIS), total radiated power (TRP), and total isotropic sensitivity (TIS), etc.; 2) develop sub-6GHz MIMO and mm-wave MIMO experimental platforms; 3) develop algorithms to retrieve antenna far-field radiation patterns in non-ideal scenarios, e.g. near-field setup with a short measurement distance, non-ideal anechoic setup with presence of unwanted reflection(s); and 4) develop cost-effective and time-efficient RC based method for TRP measurements of mm-wave massive MIMO base stations.

### A. Wireless cable solution

#### 1) Sub-6 GHz high-order MIMO radios

The conventional cabled solution is dominant in the testing industry due to its simplicity and accessible antenna ports, where testing signal can be guided to the respective antenna ports via RF cables, with balanced signal transmission and no cross-talk to the unintended antenna ports. Wireless cable solution aims to replace the cable connection functionality over-the-air, without physical RF cable connections [10]. One key is to determine the transfer matrix among the probe antenna ports and the device under test (DUT) antenna ports, and calibrate the transfer matrix. In [11], a solution is developed to estimate the transfer matrix based on reference signal received power (RSRP) values reported per DUT antenna port for 2 × 2 MIMO radios. The wireless cable testing solution is later extended to high-order MIMO terminals in [12] with a closed form calibration solution. Although the proposed approach is highly attractive, one key issue with the state-of-art wireless cable implementation is the excessive number of CE resources required in terms of digital fading channels and output interface ports. A cost-effective wireless cable implementation is proposed in the MEWS project [13] and shown in Figure 1, where a standalone phase and amplitude control matrix is introduced to reduce the CE resource. With the proposed scheme, one can achieve equivalent CE resource as the conventional cabled setup, which is highly attractive for high-order MIMO at sub-6 GHz.

#### 2) mm-wave 2×2 MIMO radios

For mm-wave mobile handsets, the antenna designs are widely different from those used for sub-6GHz systems. The mm-wave DUT antenna patterns are directive and adaptive. It is defined in 3GPP Release 15 that for a 2 × 2 MIMO system, spatial multiplexing is achieved using the polarization domain, i.e., one data stream per orthogonal polarization. For sub-6GHz MIMO radio, as discussed, one could estimate the transfer function matrix A and then calibrate it out by implementing the inverse of A. For the mm-wave band, a straightforward approach to achieve wireless cable connection is to design the transfer function matrix A such that $|A| = I$ can be directly approximated in the multi-probe setup. Therefore, there will be no need to calibrate it out. As illustrated in Figure 2 and detailed in [14], one can exploit the antenna polarization and pattern discrimination to easily achieve the wireless cable connection for the mm-wave bands, where an isolation of more than 30 dB can be achieved in the measurement setup.

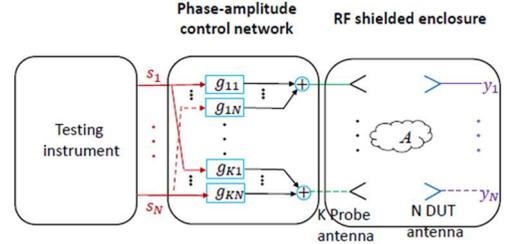

Figure 1. A cost-effective wireless cable implementation for high-order MIMO radios [13].

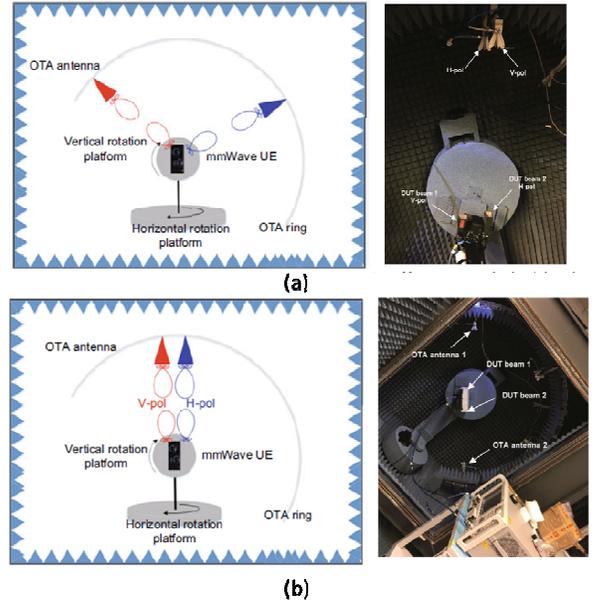

Figure 2. Achieving wireless cable connection via antenna pattern discrimination (a) and antenna polarization discrimination (b) [14].

## B. Multi-probe anechoic chamber (MPAC) method

There are a few challenges for test zone validation for MPAC solution at mm-wave bands. In the standardization, a Biconical antenna is typically used for the test zone calibration purpose, which is low gain and vertically polarized. Biconical antennas have Omni-directional pattern in the azimuth plane, yet very narrow beam width in the elevation plane. Moreover, horizontal polarized Omni-directional antenna is scarce due to technical difficulties in its realization so that validation of channel cross polarization ratio cannot be supported. In this project, one proposes to use directive antennas for the test zone validation purpose to tackle the challenges introduced by the Biconical antennas. More results to demonstrate the effectiveness of the proposed validation solution will be provided during the project.

## C. Reverberation chamber plus channel emulator (RC+CE) method

Referring to 3GPP TR 38.827, a multi-path fading propagation model (defined in 3GPP TR 38.901) needs to be used in OTA performance tests, with specified cluster parameters in spatial, Doppler, delay, and polarization domains. The RC chamber will distort the radio channels implemented using the channel emulator. Therefore, one should correct the distortion introduced by the RC chamber. Furthermore, keyhole effect (double Rayleigh) influences the statistical distribution of the generated channel. In the project, one aims to address these challenges introduced by the RC. A recent work presented in [15] shows a promising two-step closed-loop RC channel cancellation method, which is demonstrated to be rather effective. Figure 3 shows the effectiveness of the proposed method on the cancellation of the RC continuous exponential decay distribution.

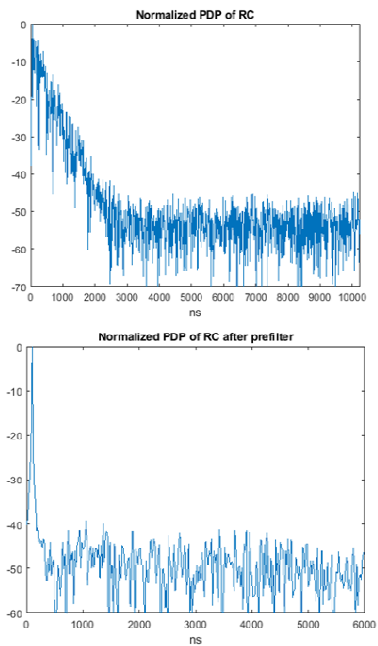

Figure 3. Channel impulse response of a single tap in the RC before (top) and after applying the cancellation algorithm [15].

## III. SUB-THZ RADIO PROPAGATION METROLOGY

This section shows the activities related to the characterization of the complex real-world sub-THz wide bandwidth antenna and radio propagation channel [16]-[17] for the planning of the short to medium range wireless communication and active services operating up to 750 GHz for the practical realization of the 5G/6G communications. The key objectives of this work is to: 1) develop traceable methodology; and 2) characterise the radio propagation channel for wide-bandwidth communications with respect to complexity, time efficiency and cost-effectiveness. This includes characterisation of the radio propagation channel up to sub-THz frequencies, i.e. up to 750 GHz with the focus on how this could feed into standardisation efforts across the spectrum, specifically in the emerging sub-THz frequency bands. The trend of considering higher frequency bands due to general availability of large unutilized bandwidths continues today into the sub-THz domain. The properties of the radio wave propagation for wide bandwidth communication channels as part of beyond 5G wireless communications are not extensively studied or experimentally verified in the open literature and therefore not incorporated in various standardisation efforts led globally by ITU-R, 3GPP, ETSI, and IEEE.

The planned tasks and activities will aim to produce original experimental and numerical techniques applied in radio channel characterisation mainly in the sub-THz bands and provide propagation data required for the planning of short/medium-range wireless communications and active services operating up to 750 GHz. The work presented here and follow-up activities will provide guidelines on validating channel sounders for sub-THz bands, characterising propagation channels for typical reflection-less and multipath-rich indoor and outdoor environments and finally establishing the relevant channel models and parameters based on massive numerically and experimentally acquired empirical datasets. The technical work enables insight and understanding of the operational parameters related to the wideband communication channels in the sub-THz frequency range paving the way for 6G applications and beyond. In order to ensure comprehensive and accurate channel models and characteristics later applied in system-level modelling and designs, practical and reliable channel sounding techniques operation up to sub-THz need to be investigated and established. There are very limited short-range controlled laboratory-based test scenarios; however, they do not present the overall picture required for large-scale implementation and application. The project consortium aims to develop and validate a traceable vector network analyser (VNA)-based channel sounding and frequency extender technology for frequencies up to 750 GHz. Figure 4 shows a schematic diagram of the VNA-based channel sounder operating at 300 GHz [18]. The developed channel sounder is limited in terms of range due to high attenuation at these frequencies and the substantial error level added due to phase errors and other noise factors inherited within the coaxial cable setup.

In practice, coaxial cable losses restrict the practical measurement range and therefore, the use of optical fibre cables to mitigate coaxial cable loss issue and hence the extension of measurement range is investigated and Figure 5 presents the details of the optical-fibre based phase-compensated sounder developed in [19]. This addresses the issue relevant to optical fibre cables being inherently sensitive to phase changes due to bending and temperature, therefore, phase-compensation scheme is required to obtain coherent phase measurements, which are pivotal for obtaining the accurate spatial profile of the propagation channel metrics. This also would allow us to extend the measurement range to up 600 m and potentially beyond. Figure 6 demonstrates the normalised power delay profiles (PDPs) obtained from the VNA-based optical fibre supported channel sounder at 28 GHz and 300 GHz [20] for comparison and to show the stability of the measurement technique and procedure followed with great potential to extend this to 500 GHz and ultimately to 750 GHz.

The next steps in this work would be to apply the validated channel sounding testbeds to conduct measurement campaigns for frequencies up to 750 GHz for various practical real-world scenarios and measurement settings. The capability to enable channel sounding up to 600 m would enable consideration for more practical use cases such as medium-range propagation channel characterisations for reflection-less and multipath-rich indoor and outdoor environments. A virtual large-scale antenna array concept is applied to enable the capture of channel parameters for variable beam width and spatial configurations [21]. The antenna pattern is measured in advance in an anechoic chamber laboratory environment and then de-embedding techniques are applied from the captured radio channels. Figure 7 presents the 100 GHz delay profile of a typical indoor corridor environment applying the virtual antenna array and showing clear channel paths and features highlighting the benefits and strong potential for the virtual array concept in further measurements and characterisations.

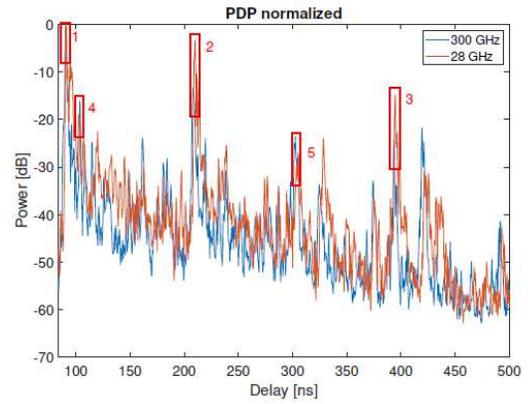

Figure 6. Normalised PDP of a long-range channel sounding experiments at 28 GHz and 300 GHz incorporating the optical fibre solution and phase-error compensation techniques. 300-GHz VNA-based phase-compensated channel sounder with optical fibre solutions [20].

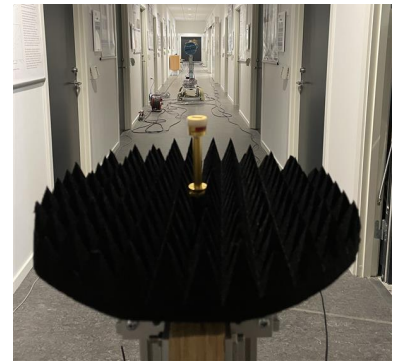

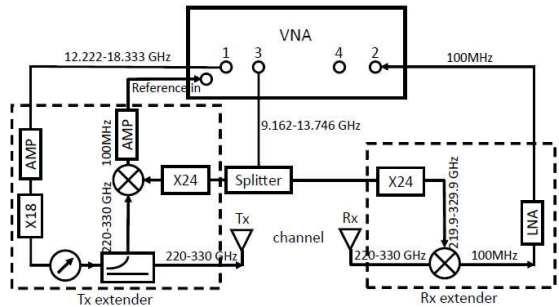

Figure 4. 300-GHz VNA-based channel sounder [18].

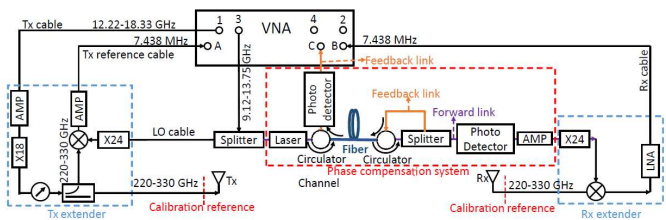

Figure 5. 300-GHz VNA-based phase-compensated channel sounder with optical fibre solutions [19].

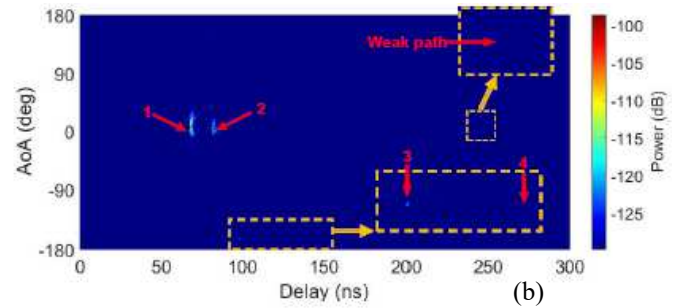

Figure 7. Virtual Antenna Array setup and concept applied for sub-THz channel sounding allowing for further flexibility and confidence in channel sounding at various environments [21]: (a) Virtual Antenna Array Set-up in a long corridor environment; (b) The Angle of Arrival delay profile clealy showing strong channel features and paths.

IV. NEW RADIO EXPOSURE METROLOGY

This section shows the activities related to the development of the measurement protocols to quantify RF exposure levels from 5G base stations by measuring power density (PD) and from mobile phones by measuring specific absorption rate (SAR) and absorbed power density (APD). These activities can be categorized in four different tasks: 1) development of accurate measurement technique for demodulation of NR signals; 2) development of testbed for traceable calibration of NR RF exposure measurement systems; 3) development and evaluation of a methodology for measuring PD exposure levels around 5G NR base stations; 4) development and evaluation of the measurement methodology for the SAR and APD of 5G NR mobile phones. Tasks 1 and 2 focus on providing technical background and know-how on the demodulation of NR signals so that a demodulation algorithm can be designed and implemented by the participating laboratories. Moreover, SI (International System of Units) traceability of the conducted, and over-the-air measurements and the demodulation of 5G NR signals are also aimed in these tasks. On the other hand, Tasks 3 and 4 focus on the development of the methodologies to quantify the NR radiation and to increase the accuracy and the measurement speed of the existing technologies in terms of PD, SAR and APD as well as their SI traceability.

For Task 1, a comprehensive review of the ETSI TS 138 211 5G NR standard was made [22] and the corresponding documentation is prepared in the form of a handbook in order to determine the key parameters for the identification of the RF power of the 5G NR resource grid allocation. This document includes also relevant information from ETSI TS 138 104, 138 213, 138 101-1, 138 212 and 138 331 standards [23]-[27]. The aim of this document is to pave way to the creation of an algorithm allowing the detection of the Primary Synchronisation Signal (PSS), Secondary Synchronisation Signal (SSS) and Demodulation Reference Signal (DM-RS), giving information on the resource allocation and bandwidth usage. A sample output of the designed detection algorithm is given in Figure 8 for the detection of PSS and SSS. In further steps in this task, in addition to synchronization signals, the SI traceability of the code selective measurement of the 5G NR signals is to be achieved for the whole resource grid. The algorithm which is currently being designed has the advantage of being fully transparent with respect to the other solutions available on different platforms. This is a crucial benefit for the correct determination of all components of the measurement uncertainty budget. The experiments for Task 1 are performed on a sample testbed. This is the precursor setup for establishing the SI traceability aimed in Task 2. After the completion of Task 1, the research will continue with the conducted and on-the-air measurements, which is aimed to have measurement uncertainties of 0.05 dB and 0.5 dB, respectively.

In the scope of the research in Task 3, the RF exposure around the 5G NR base stations is to be measured. As the number of possible cases including different traffic conditions, channel bandwidths, the radiating antenna types, etc. are too high, first efforts are currently given to the definition of the measurement scenarios and protocol which require affordable resources, but would still yield in comprehensive results to obtain the insights on the exposure characteristics. In Task 4, the main research topic remains as 5G NR RF exposure, but this time of the mobile phones in terms of SAR (sub-6GHz) and APD (mm-wave band). The efforts are given on the successful reconstruction of SAR and APD around the human phantom based on numerical simulations using surface equivalence theorem (See Figure 9). For the measurements, non-invasive methods are selected as it enables the accelerated SAR and APD acquisition due to faster surface measurements in contrast to volumetric scanning in invasive method. For different cases and frequency bands, the simulations based on proposed methods and the measurements are compared. The aim of this comparison is to minimize the error rate between simulations and measurements by improving the proposed method.

For the correct characterization of APD in mm-wave band, near field probes are designed. The linearity characteristics of these probes are also investigated in Task 4. Equivalent circuit model of near-field probes as well as their dynamic response are currently being studied. A machine learning model is also developed with the goal of finding linearization parameters per each probe based on signal and probe-specific equivalent circuit. The training of this model is currently ongoing. On the other hand, an APD calibration concept using waveguides and liquid human phantom model is developed (See Figure 10). It is implementation in 24 GHz – 36 GHz is accomplished and the preliminary uncertainty budget is calculated. An APD reflection coefficient validation concept is also successfully developed and very promising initial results for the reflection coefficient validation in 27 GHz are obtained.

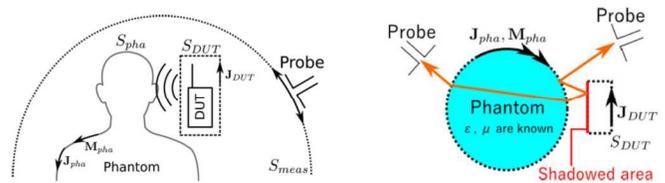

Figure 9. Reconstruction of SAR around human phantom model.

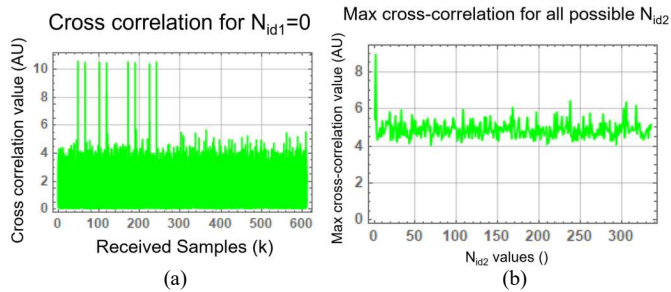

Figure 8. A sample scenario with 8 SSB Bursts for Cell ID 3: (a) PSS Detection ($N_{id1} = 0$); (b) SSS Detection ($N_{id2} = 1$).

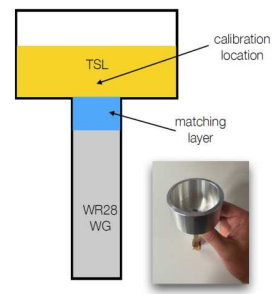

Figure 10. APD calibration concept (TSL: Tissue Simulating Liquid, WG: Waveguide).

## V. Conclusions

Metrology has a pivotal role in ensuring quality of products and end-users' confidence. This paper gives an update on the recent research and development activities from the newly funded EU Joint Research Project – MEWS in support of the metrological development for emerging wireless standards. The project addresses the emerging wireless products that underpin the 5G and 6G of mobile technologies, and aims to support the relevant standardization work by developing practical and efficient traceable measurement methods from which the standards under development can be built. Each of the described work packages within the project aims to deliver novel and original contributions that will meet the needs and requirements of future wireless standards and systems with adaptive and scalable specifications. All these developments and the corresponding results are to be incorporated into future standards through the relevant standard committees. This will have direct outcomes on wireless communications and electronics industries by ensuring the product quality and the end-user confidence.

## Acknowledgement

The project (21NRM03 MEWS) has received funding from the European Partnership on Metrology, co-financed from the European Union's Horizon Europe Research and Innovation Programme and by the Participating States. Funded by the European Union. Views and opinions expressed are however those of the author(s) only and do not necessarily reflect those of the European Union or EURAMET. Neither the European Union nor the granting authority can be held responsible for them.